\documentclass[conference]{IEEEtran}

\usepackage{amsmath}
\usepackage{amssymb}
\usepackage{graphicx}
\usepackage{xcolor}
\usepackage{array,tabularx,booktabs}
\usepackage{multirow}
\usepackage{algorithm}
\usepackage{algpseudocode}
\usepackage{tikz}
\usetikzlibrary{arrows.meta,positioning,calc,fit,backgrounds,shapes.geometric}
\usepackage{CJKutf8}

\newcolumntype{Y}{>{\raggedright\arraybackslash}X}
\definecolor{hrInk}{HTML}{14171B}
\definecolor{hrSlate}{HTML}{5C6672}
\definecolor{hrLine}{HTML}{C3C9D0}
\definecolor{hrPanel}{HTML}{FCFCFD}
\definecolor{hrWell}{HTML}{F1F3F5}
\definecolor{hrGreyF}{HTML}{EDEFF2}

\definecolor{hrBlue}{HTML}{2A5D8F}
\definecolor{hrBlueF}{HTML}{E7EEF6}
\definecolor{hrPlum}{HTML}{5E4B8B}
\definecolor{hrPlumF}{HTML}{EEEAF5}
\definecolor{hrGreen}{HTML}{226B54}
\definecolor{hrGreenF}{HTML}{E4F0EA}
\definecolor{hrAmber}{HTML}{C96A2A}
\definecolor{hrAmberF}{HTML}{F9E1CE}

\providecommand{\hrnohyph}{\hyphenpenalty=10000\exhyphenpenalty=10000\relax}
\providecommand{\hrtiny}{\fontsize{6}{6.6}\selectfont}
\providecommand{\hrsmall}{\fontsize{7}{8}\selectfont}

\tikzset{
  hrfig/.style={x=1.0547cm, y=1.0547cm, font=\hrsmall, line width=0.5pt,
    text=hrInk, >={Stealth[round,length=3.4pt,width=2.6pt]}},
  hrpanel/.style={draw=hrLine, fill=hrPanel, rounded corners=2.6pt, line width=0.5pt},
  hrwell/.style={draw=hrLine, fill=hrWell, rounded corners=1.6pt, line width=0.35pt},
  hrcard/.style={draw=hrLine, fill=white, rounded corners=2pt, line width=0.4pt},
  hrcardB/.style={hrcard, draw=hrBlue!55, fill=hrBlueF},
  hrcardP/.style={hrcard, draw=hrPlum!55, fill=hrPlumF},
  hrcardG/.style={hrcard, draw=hrGreen!55, fill=hrGreenF},
  hrtitle/.style={font=\footnotesize\bfseries, inner sep=0pt, anchor=west},
  hrsub/.style={font=\hrtiny, text=hrSlate, inner sep=0pt, anchor=west},
  hrlab/.style={font=\hrtiny, text=hrInk, inner sep=0pt},
  hrhead/.style={font=\hrsmall\bfseries, inner sep=0pt, anchor=west},
  hrchip/.style={rounded corners=1.6pt, inner xsep=2.6pt, inner ysep=1.6pt,
    text height=1.45ex, text depth=0.35ex,
    font=\hrtiny, line width=0.4pt, execute at begin node=\hrnohyph},
  hrchipB/.style={hrchip, draw=hrBlue,  fill=hrBlueF,  text=hrBlue!72!black},
  hrchipP/.style={hrchip, draw=hrPlum,  fill=hrPlumF,  text=hrPlum!72!black},
  hrchipG/.style={hrchip, draw=hrGreen, fill=hrGreenF, text=hrGreen!62!black},
  hrchipA/.style={hrchip, draw=hrAmber, fill=hrAmberF, text=hrAmber!70!black},
  hrchipN/.style={hrchip, draw=hrLine,  fill=hrGreyF,  text=hrSlate},
  hrflow/.style={->, draw=hrSlate, line width=0.6pt},
  hrflowB/.style={hrflow, draw=hrBlue},
  hrflowP/.style={hrflow, draw=hrPlum},
  hrflowG/.style={hrflow, draw=hrGreen},
  hrbus/.style={draw=hrSlate, line width=0.6pt},
  hrsel/.style={draw=hrSlate, line width=0.5pt, dash pattern=on 1.6pt off 1.2pt},
  hrhair/.style={draw=hrLine, line width=0.4pt},
  hrtext/.style={draw=hrSlate!50, line width=1.1pt, line cap=round},
}

\usepackage{pgfplots}
\pgfplotsset{compat=1.18}
\usepackage{url}
\usepackage[hidelinks]{hyperref}

\makeatletter
\def\blfootnote{\xdef\@thefnmark{}\@footnotetext}
\makeatother

\begin{document}

\title{HistoRAG: A Citation-Grounded Question Answering Assistant for Teaching with Scanned Local History and Heritage Archives}

\author{\IEEEauthorblockN{Hongzhou Duan\textsuperscript{1}, Zikun Guo\textsuperscript{2,3,\dag}}
\IEEEauthorblockA{\textsuperscript{1}Mudanjiang Normal University, Heilongjiang Province, China\\ m18245337405@163.com}
\IEEEauthorblockA{\textsuperscript{2}School of Autonomous Driving, Xinwei Institute of Artificial Intelligence, 572022, Sanya, Hainan, China}
\IEEEauthorblockA{\textsuperscript{3}Center for Mathematical Research, Xinwei Institute of Artificial Intelligence, 572022, Sanya, Hainan, China\\ gzk798412226@gmail.com}}

\maketitle
\bstctlcite{IEEEexample:BSTcontrol}
\blfootnote{\dag\,Corresponding author: Zikun Guo (gzk798412226@gmail.com).}

\begin{abstract}
Teachers who prepare lessons on local history and cultural heritage work from material that is hard to use. The primary sources are scanned books without a text layer, and the supporting records are administrative catalogs released as spreadsheets. A general chatbot answers such questions fluently but without a verifiable source, which is the property a teacher needs most. This paper presents HistoRAG, a question answering assistant that answers from one regional collection and cites a volume and a page for every fact. HistoRAG transcribes each page with a vision language model and keeps a line level confidence from the token probabilities. It builds three stores from the same collection: a hybrid text index, a relational catalog database, and a knowledge graph extracted only from entity dense passages. A lightweight router sends each question to the stores it needs, so that counting questions reach the database and relational questions reach the graph. We build a benchmark of 516 questions over a collection of 36 scanned volumes and the heritage catalogs of the same region, covering statistical, factual, temporal, and multi-hop questions. HistoRAG answers more questions correctly than passage retrieval baselines and than a graph based retrieval system, at a far smaller cost per question. The assistant runs behind a chat interface, so a teacher can check any statement against the page it came from.
\end{abstract}


\section{Introduction}
\label{sec:intro}

Lessons on local history and cultural heritage draw on two kinds of source. The narrative record consists of chronicles, organizational histories and biographies, which exist only as scanned books without a text layer. The administrative record consists of official catalogs of protected sites and museum objects, released as spreadsheets and word processor tables. Finding one fact means turning pages in several volumes, and checking one number means reading a table never designed for reading.

A general purpose chatbot answers from parameters rather than from the collection at hand, and gives no source that a teacher can check. Surveys of educational applications state the requirement directly: retrieval grounding is what makes a generated answer usable in teaching, because the material behind it can be inspected \cite{li2025ragedu}. An expert rated comparison of two tutors found that retrieval improved instructional quality, while unsupported statements still appeared \cite{adenuga2026ragtutor,zhou2024trustrag}.

Grounding an assistant in a scanned regional collection raises three technical problems. First, the text comes from optical character recognition, and recognition errors propagate into retrieval and generation \cite{zhang2025ohrbench}. A low character error rate does not guarantee a good answer \cite{sun2026goodocr}. Second, counting and filtering questions over catalogs are poorly served by passage retrieval, because the answer is an aggregate over many rows. Third, graph based retrieval improves relational and multi-hop questions \cite{edge2024graphrag,han2025ragvsgraphrag}, but extracting a graph from a whole collection with a language model is expensive.

\begin{figure*}[t]
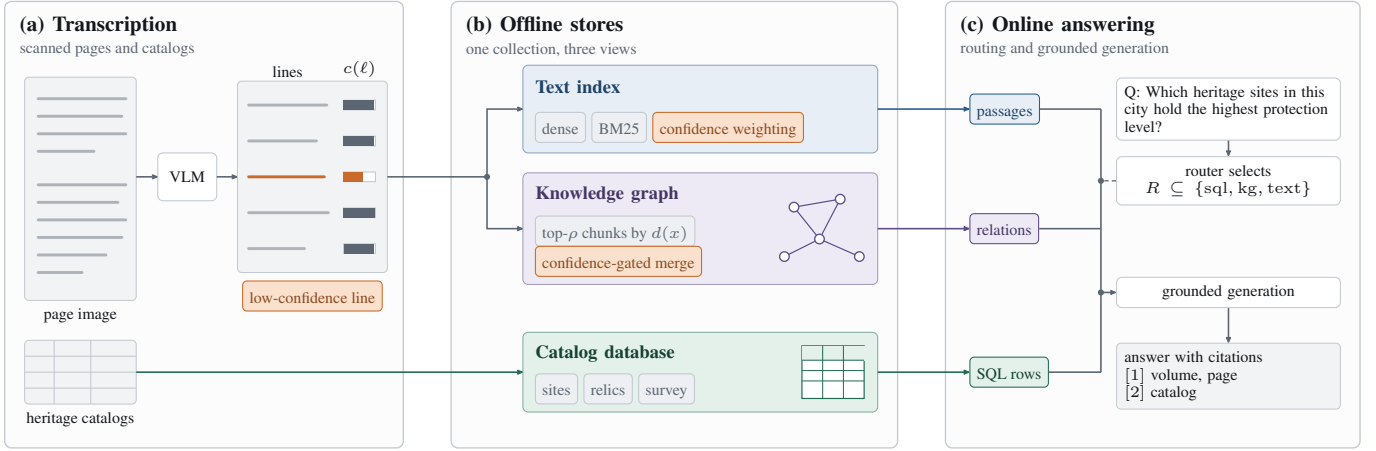

  \centering\noindent
\begin{tikzpicture}[hrfig]
  \def\pAx{0}
  \def\pBx{5.6}
  \def\pCx{11.8}
  \def\pRowy{0}

  \begin{scope}[shift={(\pAx,\pRowy)}]\input{figs/fig1/panel_a.tex}\end{scope}
  \begin{scope}[shift={(\pBx,\pRowy)}]\input{figs/fig1/panel_b.tex}\end{scope}
  \begin{scope}[shift={(\pCx,\pRowy)}]\input{figs/fig1/panel_c.tex}\end{scope}

  \coordinate (abus) at (6.05,0);

  \draw[hrflow] (aOutText) -- (aOutText -| abus) |- (bIn1);
  \draw[hrflow] (aOutText -| abus) |- (bIn2);
  \fill[hrSlate] (aOutText -| abus) circle (1.15pt);

  \draw[hrflowG] (aOutCat) -- (bIn3);

  \draw[hrflowB] (bOut1) -- (cIn1);
  \draw[hrflowP] (bOut2) -- (cIn2);
  \draw[hrflowG] (bOut3) -- (cIn3);
\end{tikzpicture}
  \caption{Overview of HistoRAG. (a) Each scanned page is transcribed by a vision language model, and every output line keeps a confidence $c(\ell)$ from its token probabilities. (b) The collection is organized into a text index with confidence-weighted rank fusion, a knowledge graph extracted from entity dense chunks with confidence-gated entity merging, and a database normalized from heritage catalogs. (c) A router selects the stores for each question, and the generator cites a volume and page, or a catalog, for every fact.}
  \label{fig:overview}
\end{figure*}

We present HistoRAG, a question answering assistant for teaching with such collections (Figure~\ref{fig:overview}). Each page is transcribed by a vision language model, and each output line keeps a confidence from its token probabilities. The collection is then organized into three stores: a hybrid dense and lexical index for narrative questions, a relational database normalized from heterogeneous catalogs for counting and listing questions, and a knowledge graph for relational and multi-hop questions, extracted only from passages with high entity density. A lightweight router selects the stores per question, and the generator cites the source page of every fact.

The main contributions of this paper are summarized as follows:
\begin{itemize}
\item We present a teaching assistant that answers from one scanned regional collection and attaches a volume and page to every fact, so that a teacher can verify a statement before using it in class.

\item We formulate the task as unified retrieval over narrative text, structured catalogs, and an entity graph built from the same collection, and we route each question to the stores it needs.

\item We propose selective graph construction, which extracts entities and relations only from passages ranked by a rule based entity density score. The retained fraction controls the trade-off between indexing cost and answer accuracy.

\item We build a benchmark of 516 questions over 36 scanned volumes and the heritage catalogs of the same region, covering statistical, factual, temporal, and multi-hop questions, and we compare HistoRAG with passage retrieval and graph retrieval baselines.
\end{itemize}

The rest of the paper is organized as follows. Section~\ref{sec:related} reviews related work. Section~\ref{sec:method} presents the framework. Section~\ref{sec:exp} describes the benchmark and the experimental setup, and reports the results. Section~\ref{sec:conclusion} concludes the paper.

\section{Related Work}
\label{sec:related}

\paragraph{RAG in education}
Retrieval grounding is the property that makes generated text usable in teaching, and a systematic survey of educational applications organizes the field around it \cite{li2025ragedu}. An expert rated comparison of an algebra tutor with and without retrieval reports better instructional quality with retrieval, together with unsupported statements that call for evidence governance \cite{adenuga2026ragtutor}. Surveys of trustworthiness in retrieval augmented generation reach the same requirement in the general setting \cite{zhou2024trustrag}. These studies evaluate tutoring dialogue with curricular material. HistoRAG addresses the preparation stage instead, where the material is a scanned regional collection and the teacher needs a page reference for each fact.

\paragraph{OCR noise in RAG}
Document parsers \cite{wang2024mineru} and vision language models \cite{bai2025qwen3vl} convert page images into text for RAG, and the same backbones serve other recognition tasks \cite{guo2025crop}. OHRBench \cite{zhang2025ohrbench} shows that semantic and formatting noise both degrade retrieval and generation, and a later benchmark reports retrieval failures even at a low word error rate \cite{sun2026goodocr}. AncientDoc \cite{yu2025ancientdoc} evaluates transcription and knowledge questions on historical books. These studies measure the effect of noise. HistoRAG keeps a confidence signal from recognition and uses it during retrieval and entity normalization.

\paragraph{Graph based RAG}
GraphRAG \cite{edge2024graphrag} extracts an entity graph with an LLM and summarizes graph communities for global questions. Surveys and systematic comparisons show that graphs help relational and multi-hop questions, while plain RAG remains strong on detail questions \cite{peng2024graphragsurvey,han2025ragvsgraphrag,xiang2025whengraphs}. Much of the recent work reduces cost. LightRAG \cite{guo2024lightrag} uses dual level keyword retrieval over a lighter graph. HippoRAG and HippoRAG 2 \cite{gutierrez2024hipporag,gutierrez2025hipporag2} apply personalized PageRank over a passage and phrase graph. LiteRAG \cite{colltejeda2026literag} replaces LLM control at retrieval time with algorithmic exploration. These methods still extract from every passage. HistoRAG reduces the extraction cost itself by selecting passages before any LLM call.

\paragraph{Heterogeneous source QA}
Quasar \cite{christmann2024quasar} answers over text, tables, and knowledge bases in one pipeline, ER-RAG \cite{xia2025errag} models heterogeneous sources with an entity relationship schema, and HetaRAG \cite{yan2025hetarag} spans vector, graph, full text, and relational stores. RouteRAG \cite{guo2025routerag} learns a routing policy with reinforcement learning, and KAG \cite{liang2024kag} combines graph reasoning with vector retrieval. DIN-SQL \cite{pourreza2023dinsql} decomposes text-to-SQL generation with self-correction. HistoRAG routes with a prompt and no training, and grounds SQL generation in the value sets of the catalog tables.

\paragraph{Reliable and efficient models}
Models can follow misleading user input even when it contradicts the evidence, and filtering non-evidential cues reduces this failure \cite{zhou2026flattery,xu2025medsyco}. Causal auditing reveals when a model relies on spurious inputs instead of the evidence \cite{guo2026cadet}. This motivates the requirement that every generated fact cites retrieved evidence. On efficiency, model compression and lightweight architectures keep accuracy at a small budget \cite{guo2023idsextract,guo2025cat,fan2025tsdcaba}. HistoRAG pursues the same goal at the system level, by limiting LLM calls.

\paragraph{Historical archive QA}
Graph RAG has been applied to classical historical texts with a relation extraction model \cite{yang2025historicalgraphrag}, and to handwritten records with an agent that writes graph queries and expands literals by word spotting \cite{nicolau2026robust}. That graph is built with human assistance. LLMs have also been combined with visual data for place-level understanding \cite{han2026street}. HistoRAG targets printed books of the twentieth century and builds all three stores automatically.

\section{Method}
\label{sec:method}

\subsection{Problem Setting}

The input collection has two parts. The first is a set of scanned volumes $\mathcal{V}$, where each volume is a sequence of page images. The second is a set of catalog files $\mathcal{T}$, each holding one or more tables of heritage records. Given a natural language question $q$, the system returns an answer $a$ and a list of citations. Each citation points to a volume and page, or to a catalog.

HistoRAG has an offline stage and an online stage. The offline stage builds three stores: a text index $\mathcal{I}$, a relational database $\mathcal{D}$, and a knowledge graph $\mathcal{G}$. The online stage routes $q$ to a subset of these stores and generates $a$ from the retrieved evidence.

\subsection{Confidence-Carrying Transcription}

Each page image is transcribed by a vision language model \cite{bai2025qwen3vl}. The prompt asks for the body text in reading order, with captions and footnotes kept and running headers removed. Decoding is greedy, and the model returns the log probability of every output token.

Let a transcribed line $\ell$ consist of tokens $t_1,\dots,t_n$. We define the line confidence and the minimum token confidence as
\begin{equation}
c(\ell)=\exp\Big(\frac{1}{n}\sum_{i=1}^{n}\log p(t_i)\Big), \qquad
c_{\min}(\ell)=\min_{i} p(t_i).
\end{equation}
Pages whose only image is a placeholder are marked as missing and skipped.

\subsection{Text Index}

Pages are cleaned before chunking. Short lines repeating at the top or bottom of many pages are removed as running headers, together with page numbers and table of contents entries. A paragraph that ends a page without closing punctuation is joined with the first paragraph of the next page. Paragraphs are packed into chunks of about 500 characters, and longer paragraphs are split at sentence boundaries.

Each chunk $x$ keeps its page range and a confidence $\bar{c}(x)$, the mean of $c(\ell)$ over its lines. The index stores a dense embedding of each chunk \cite{chen2024bgem3} and a BM25 representation over word segmented text.

\subsection{Catalog Database}

Heritage catalogs differ in layout: headers may span two rows, region names may appear as section rows, and one list may be published twice in different formats. We map each table to a common schema by matching header cells against synonym lists for fields such as name, region, city, county, protection level, and collecting institution. Two row headers are resolved by the column positions of the second row, and records are deduplicated on name, city, county, and institution.

The result is three relational tables: immovable sites, movable objects, and a regional survey of people, events, and memorials.

\subsection{Selective Graph Construction}

Extracting a graph from every chunk is the most expensive offline step. We first rank chunks by a rule based entity density
\begin{equation}
d(x)=\frac{100}{|x|}\big(n_{\mathrm{ne}}(x)+n_{\mathrm{unit}}(x)+n_{\mathrm{date}}(x)\big),
\end{equation}
where $|x|$ is the chunk length in characters. The term $n_{\mathrm{ne}}$ counts person, place, and organization words from a part of speech tagger. The terms $n_{\mathrm{unit}}$ and $n_{\mathrm{date}}$ count organizational designations and dates matched by patterns. Only the top fraction $\rho$ of chunks by $d(x)$ is sent to the LLM.

For each selected chunk, the LLM returns entities and relations under a JSON schema. Entity types are person, organizational unit, organization, place, event, site, and document. Relation types include holding a position, belonging to a unit, participating in an event, leading, dying at, and alias. The prompt asks the model to copy suspected recognition errors as written, so that normalization can handle them in one place.

\subsection{Confidence-Gated Entity Normalization}

Mentions are normalized in three steps. Surface forms are cleaned, and honorific suffixes are removed from person names. The second step merges a mention with a name listed as its alias, unless that name is itself a frequent independent entity, which happens when the extraction model reports two distinct people as aliases of each other. The third step merges recognition variants.

Let $u$ and $v$ be two surface forms of the same type and length, with mention counts $f(u) \le f(v)$ and edit distance one. Let $L(u)$ be the set of transcribed lines in which $u$ occurs. We merge $u$ into $v$ only if
\begin{equation}
\max_{\ell \in L(u)} c(\ell) < \tau
\quad \text{and} \quad
f(v) \ge \kappa\, f(u),
\end{equation}
where $\tau$ is a confidence threshold and $\kappa$ is a frequency ratio. The first condition requires every occurrence of the rare form to come from a line that the model itself read with low confidence. A name that appears on one clearly read line is therefore kept, even when a similar frequent name exists.

Three further conditions block merges that the edit distance alone would allow. Names shorter than three characters are excluded, since one character is half of such a name. Unit designations and document titles are excluded, since they differ by numerals that carry meaning. Person names with different surnames are excluded. Using the confidence of the enclosing chunk instead of the line makes the first condition vacuous, because a chunk of several hundred characters almost always contains one low probability token.

\subsection{Routing and Retrieval}

At query time, the router prompts the LLM to select a subset $R \subseteq \{\mathrm{sql}, \mathrm{kg}, \mathrm{text}\}$ and to list the entity names in $q$. Statistical and listing questions go to the database, relational questions about named entities go to the graph, and narrative questions go to the text index. The design splits a heterogeneous workload into homogeneous parts, each served by a dedicated module \cite{guo2025cag}, and store selection is itself a task allocation problem, studied with coevolutionary search, clustering, structural priors, and reinforcement learning \cite{adedigba2025ccga,adedigba2025ivec,guo2026structural,guo2025dyt}.

\paragraph{Database route}
The LLM writes a read-only SQL query from the table schema. The prompt also lists the distinct source catalogs and the most frequent protection levels. Catalog categories are recorded only in these values, so the value list lets the model filter by the right field. A query that fails to execute is returned to the model with the error message for one retry.

\paragraph{Graph route}
Each entity name from the router is linked to a graph node by exact name, alias, and then substring match. The route returns the relations within $h$ hops of the linked nodes, each with the page of its source chunk.

\paragraph{Text route}
Dense and BM25 retrieval each return a ranked list. The two lists are merged by reciprocal rank fusion, and the fused score is weighted by chunk confidence:
\begin{equation}
s(x)=\Big(\sum_{r \in \{\mathrm{dense},\,\mathrm{bm25}\}} \frac{1}{k + \mathrm{rank}_r(x)}\Big)\cdot\big(\alpha+(1-\alpha)\,\bar{c}(x)^{\gamma}\big).
\end{equation}
The weight lowers the rank of poorly recognized chunks without removing them. If the selected routes return no evidence, the text route is used as a fallback.

\subsection{Grounded Generation}

The evidence from all routes is numbered and passed to the LLM with the question. The prompt requires a citation mark after every factual statement, a separate account when sources disagree, and an explicit statement when the evidence is insufficient. The system appends the source list, with volume title and page, or the catalog name for database results.

\section{Experiments}
\label{sec:exp}


\subsection{Corpus}

The corpus is one regional collection on twentieth century local history and cultural heritage. It contains 36 scanned volumes with 13,970 pages: organizational histories, biographies, name rosters, documentary compilations, and regional surveys of heritage sites. The catalog part contains official lists of immovable and movable heritage, and a survey of people, events, and memorials. After normalization, the database holds 1,943 site records, 17,385 object records, and 984 survey records.

\begin{table}[t]
\centering
\caption{Corpus and knowledge base in numbers.}
\label{tab:corpus}
\footnotesize
\begin{tabular}{lr}
\toprule
Quantity & Value \\
\midrule
Volumes / transcribed pages & 36 / 13,601 \\
Text chunks / sent to extraction & 20,554 / 8,208 \\
Graph entities / relations & 60,564 / 129,148 \\
Catalog records & 20,312 \\
Benchmark questions & 516 \\
\bottomrule

\end{tabular}
\end{table}

Table~\ref{tab:corpus} summarizes the collection and the stores built from it. Of the 13,970 scanned pages, 13,601 carry text, and the rest are blank placeholders in the source scans. With $\rho=0.4$, extraction runs on 8,208 of the 20,554 chunks.

\subsection{Benchmark}

The benchmark has four question types. Statistical questions ask for counts or lists over the catalogs, and their gold answers are computed by SQL. Factual and temporal questions are generated from single chunks of the narrative volumes. Multi-hop questions are generated from two chunks in different volumes that mention the same graph entity. Every generated question is checked by a separate model call, which must answer it from the evidence alone. A multi-hop question is kept only if neither chunk alone is sufficient. A final pass removes questions whose answer string appears in the question, questions that refer to the source text instead of naming their subject, and temporal questions whose answer is relative to an unnamed date.
The benchmark holds 516 questions: 150 statistical, 220 factual, 44 temporal, and 102 multi-hop. A hash of the question identifier assigns 166 questions to a development split and 350 to a test split. We tune all settings on the development split and report the test split.

\subsection{Baselines and Metrics}

We compare HistoRAG with an LLM without retrieval, dense RAG, BM25 RAG, and hybrid RAG, all of which read the same transcribed text. As a graph retrieval baseline we run LightRAG \cite{guo2024lightrag} on the narrative volumes of the same corpus. Every system uses the same LLM for extraction and generation, and the same embedding model for dense retrieval, so the comparison isolates the retrieval design.

Answer correctness is judged by an LLM against the gold answer, with exact match required for counts. We also report character level F1, average normalized Levenshtein similarity (ANLS), and evidence recall for questions with gold chunks. Efficiency is measured by the number of LLM tokens used to build the index and the mean tokens and latency per question.

\subsection{Implementation Details}

Transcription uses Qwen3-VL-30B-A3B \cite{bai2025qwen3vl}. Extraction, routing, SQL generation, and answer generation use Qwen3.8-27B with thinking disabled. Dense retrieval uses BGE-M3 \cite{chen2024bgem3}. Models are served with vLLM \cite{kwon2023vllm}. Unless stated otherwise, $\rho$, $\tau$, $\kappa$, $k$, $\alpha$, $\gamma$, and $h$ are set to 0.4, 0.9, 5, 60, 0.5, 4, and 1.

\subsection{Main Results}

\begin{table}[t]
\centering
\caption{Answer accuracy by question type and efficiency.}
\label{tab:main}
\footnotesize
\setlength{\tabcolsep}{2.5pt}
\begin{tabular}{lcccccc}
\toprule
Method & Stat. & Fact & Time & Multi & All & Idx.\ tok. \\
\midrule
LLM only         & 4.7 & 13.1 & 9.7 & 4.4 & 8.6 & 0.0M \\
Dense RAG        & 14.1 & 72.4 & 93.5 & 19.1 & 46.3 & 0.0M \\
BM25 RAG         & 54.7 & 91.7 & 87.1 & 29.4 & 68.0 & 0.0M \\
Hybrid RAG       & 30.2 & 87.6 & 90.3 & 27.9 & 58.9 & 0.0M \\
\midrule
HistoRAG         & 92.5 & 86.9 & 96.8 & 48.5 & 82.0 & 20.9M \\
\bottomrule

\end{tabular}
\end{table}

\begin{figure}[t]
\centering
\begin{tikzpicture}
\begin{axis}[
  ybar, width=\columnwidth, height=3.5cm, bar width=4.2pt,
  font=\footnotesize, ymin=0, ymax=105, ylabel={Accuracy (\%)},
  symbolic x coords={Stat.,Fact,Time,Multi}, xtick=data,
  ymajorgrids=true, grid style={hrLine!60, line width=0.3pt},
  axis line style={hrLine}, tick style={hrLine},
  legend style={at={(0.5,1.02)}, anchor=south, legend columns=4, draw=none,
                font=\scriptsize, column sep=4pt},
  enlarge x limits=0.18, tick label style={font=\scriptsize}]
\addplot[fill=hrSlate!45, draw=hrSlate] table[x=type, y=llm_only] {figs/acc.dat};
\addplot[fill=hrBlue!45, draw=hrBlue] table[x=type, y=bm25_rag] {figs/acc.dat};
\addplot[fill=hrPlum!45, draw=hrPlum] table[x=type, y=hybrid_rag] {figs/acc.dat};
\addplot[fill=hrGreen!65, draw=hrGreen] table[x=type, y=full] {figs/acc.dat};
\legend{LLM only, BM25, Hybrid, HistoRAG}
\end{axis}
\end{tikzpicture}
\caption{Accuracy by question type on the test split.}
\label{fig:acc}
\end{figure}
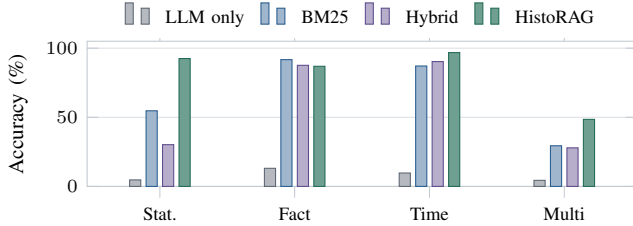

On the test split of 350 questions, HistoRAG answers 82.0\% correctly, against 68.0\% for BM25 retrieval, the strongest retrieval baseline (Table~\ref{tab:main}). The gain comes from two question types. Statistical questions rise from 54.7\% to 92.5\%, because the database route answers them with one aggregate query. Multi-hop questions rise from 29.4\% to 48.5\%, because the graph route returns the relations that connect two volumes. Factual and temporal questions are already served by passage retrieval, and the retrieval baselines stay close to HistoRAG on them.

\subsection{Contribution of Each Route}

Removing one route at a time shows where the gain comes from. Without the database route, overall accuracy falls to 64.9\%. Without the graph route, multi-hop accuracy falls from 48.5\% to 26.5\%. Querying every store without routing costs 16.1 points on multi-hop questions, because catalog rows and graph facts crowd out the passages that carry the answer.

\subsection{Selective Graph Construction}

\begin{figure}[t]
\centering
\begin{tikzpicture}
\begin{axis}[
  width=\columnwidth, height=3.4cm, font=\footnotesize,
  xlabel={Indexing tokens (millions)}, ylabel={Multi-hop acc. (\%)},
  ymajorgrids=true, xmajorgrids=true, grid style={hrLine!60, line width=0.3pt},
  axis line style={hrLine}, tick style={hrLine},
  xmin=4, xmax=23, tick label style={font=\scriptsize},
  nodes near coords={$\rho=\pgfplotspointmeta$},
  every node near coord/.append style={font=\scriptsize, yshift=2pt, text=hrSlate},
  point meta=explicit symbolic]
\addplot[mark=*, mark size=1.8pt, color=hrGreen, line width=0.8pt]
  table[x=tokens, y=acc, meta=rho] {figs/rho.dat};
\end{axis}
\end{tikzpicture}
\caption{Indexing cost against multi-hop accuracy as the retained fraction $\rho$ varies.}
\label{fig:rho}
\end{figure}
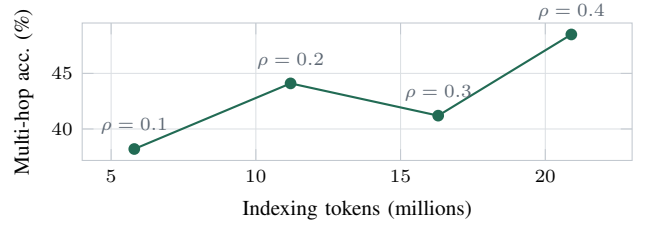

Figure~\ref{fig:rho} varies the retained fraction $\rho$ and rebuilds the graph from the same extractions. Multi-hop accuracy rises from 38.2\% at $\rho=0.1$ to 48.5\% at $\rho=0.4$. Half of the indexing budget already recovers most of that gain: $\rho=0.2$ reaches 44.1\% with 11.2 million tokens, and the remaining 4.4 points cost a further 9.7 million. The value at $\rho=0.3$ falls below this trend by two questions out of 68, which is within the variation of this subset.

\subsection{Comparison with a Graph Retrieval Baseline}

LightRAG indexes the narrative volumes of the same corpus and answers from its own graph and passages. Its questions are therefore limited to the factual, temporal, and multi-hop types, and Table~\ref{tab:lightrag} reports both systems on that matched subset of the test split.

\begin{table}[t]
\centering
\caption{HistoRAG and a graph retrieval baseline on the matched subset.}
\label{tab:lightrag}
\footnotesize
\setlength{\tabcolsep}{3.5pt}
\begin{tabular}{lcccc}
\toprule
Method & Acc. & Tok./q. & Lat.\ (s) & Idx.\ tok. \\
\midrule
LightRAG   & 70.9 & 122,008 & 41.2 & 5.7M \\
HistoRAG   & 77.5 & 4,052 & 20.6 & 20.9M \\
\bottomrule

\end{tabular}
\end{table}

On the 244 matched questions, HistoRAG answers 77.5\% correctly against 70.9\% for LightRAG, and it uses 4,052 tokens per question against 122,008. The gap in query cost follows from the retrieval design: LightRAG expands keywords over its graph and passes a large context to the generator, while routing sends most questions to one store. The index cost is not comparable in the same way, since LightRAG was built over a 2,364 chunk subset that contains every gold passage, and HistoRAG was built over the whole collection.

\subsection{The Assistant in Use}

The system is served behind a chat interface, so a teacher reaches it from a browser without installing anything. Every statement in an answer carries a bracketed mark, followed by the source list with volume title and page, or with the catalog name. When sources disagree, the generator reports the accounts separately rather than choosing one, and the teacher decides which to use in class.

\section{Conclusion}
\label{sec:conclusion}

This paper presented HistoRAG, a question answering assistant for teaching with scanned local history and heritage collections. HistoRAG carries line level recognition confidence from transcription into retrieval and entity normalization. It organizes one collection into a hybrid text index, a catalog database, and a selectively built knowledge graph, and routes each question to the stores it needs. On a benchmark of 516 questions over one regional collection, HistoRAG answers more questions correctly than passage retrieval and graph retrieval baselines (Table~\ref{tab:main}). Every answer carries a volume and page, so a teacher can check a statement before using it. The database route carries the gain on statistical questions, and the graph route carries the gain on multi-hop questions.

Future work will investigate learned routing policies, confidence estimates calibrated against manual transcriptions, and extension to handwritten and vertically typeset documents.

\section*{Acknowledgment}
This work was supported by the research project ``Construction and Development of Dynamic Semantic Application Framework'' (No. 1354MSYQN020). I sincerely thank my supervisor for the professional guidance and valuable comments throughout the research and thesis writing. I am also grateful to my teammates and colleagues for their helpful discussions and technical support during the project implementation.

\bibliographystyle{IEEEtran}
\bibliography{refs}

\end{document}